# Assessing metadata privacy in neuroimaging

**Emilie Kibsgaard[1], Anita Sue Jwa[2], Christopher J Markiewicz[2], David Rodriguez Gonzalez[3], Judith Sainz Pardo[3], Russell A. Poldrack[2] & Cyril R. Pernet[1]**

[1] Neurobiology Research Unit, Copenhagen University Hospital, Copenhagen, Denmark

[2] Advanced Computing and e-Science Group, Instituto de Física de Cantabria, Spain

[3] Department of Psychology, Stanford University, Stanford, CA, United States

corresponding author: Dr Cyril Pernet cyril.pernet@nru.dk

**Abstract**

The ethical and legal imperative to share research data without causing harm requires careful attention to privacy risks. While mounting evidence demonstrates that data sharing benefits science, legitimate concerns persist regarding the potential leakage of personal information that could lead to reidentification and subsequent harm. We reviewed metadata accompanying neuroimaging datasets from heterogeneous studies openly available on OpenNeuro, involving participants across the lifespan—from children to older adults—with and without clinical diagnoses, and including associated clinical score data. Using metaprivBIDS (https://github.com/CPernet/metaprivBIDS), a software application for BIDS-compliant tsv/json files that computes and reports different privacy metrics (k-anonymity, k-global, l-diversity, SUDA, PIF), we found that privacy is generally well maintained, with serious vulnerabilities being rare. Nonetheless, issues were identified in nearly all datasets and warrant mitigation. Notably, clinical score data (e.g., neuropsychological results) posed minimal reidentification risk, whereas demographic variables — age, sex assigned at birth, sexual orientations, race, income, and geolocation — represented the principal privacy vulnerabilities. We outline practical measures to address these risks, enabling safer data sharing practices.

## Introduction

Sharing data is beneficial in many ways: it accelerates progress in our fundamental understanding of the topic addressed by the data, it improves publication and data quality, it reduces the cost of research and increases the return on current research investments, it fosters research and advances in practices, and it is a requirement for reproducible science (Poline et al., 2012). Together, these maximise the potential benefits of the research, which is an ethical requirement for human subjects research (Brakewood & Poldrack, 2013). In neuroimaging, data collected for research often fall under some form of data protection law (e.g. GDPR in Europe or HIPAA in the USA), and appropriate measures must be taken before sharing the data to reduce the potential for reidentification. Personal data can be broadly defined as data about a particular living individual. One specific type of personal data of interest here is *sensitive personal data*. This concept of sensitive data (a.k.a special category data in GDPR) exists and is lawfully regulated in many countries, in particular, those producing most of neuroimaging data (USA, Canada, EU, Australia, China, Japan, South Korea, etc see e.g. *DLA Piper Global Data Protection Laws of the World*) and refers to data relating to sensitive areas of an individual's life, such as religion, race and political beliefs and, importantly for neuroimaging, any data relating to physical (including genetic and biometric) or mental health conditions.

In Human neuroimaging research, imaging data are typically accompanied by information about the participants. For instance, in the Brain Imaging Data Structure (Gorgolewski et al., 2016), it is recommended to have a participants.tsv file (with its .json sidecar file) that describes, at minimum, age, sex and handedness. In many cases, these files also include sensitive personal data related to physical and mental health, as this is at the heart of neuroscience research. This allows for comparisons between patients and control participants, as well as tests for brain-genetic, brain-disease, and brain-behavioural associations. Here, we present data privacy tools that quantify the potential for re-identification of individuals and apply them to open datasets. These tools operate on the principle of evaluating the degree to which personal data can be linked to a specific individual when shared or analysed. These tools do not examine the meaning of the variables; instead, they analyse the joint combinations and distributions of variables to identify individuals who are highly distinct in their combination of values. Human interpretation of variable relevance remains essential, but tools help inform data anonymisation strategies and compliance with legal and ethical standards by revealing potential issues. Here, by analysing real datasets, we show that some individuals may be at risk of re-identification based on their demographic and clinical data due to their particular group affiliation and the data sampling. We also present simple mitigating strategies that reduce risks, thus ensuring that datasets are both privacy-preserving and functionally useful for research or other applications.

## *Tools and metrics for data privacy*

There exist many different metrics of privacy, each with domain-specific properties. They have been broadly divided into methods quantifying uncertainty, information gain/loss, indistinguishability, data similarity, time, adversary's success probability, accuracy/precision, and error (Wagner & Eckhoff, 2019). Yet, state-of-the-art privacy and anonymisation software solutions for sensitive data primarily focus on tabular data. Notable tools include ARX, an open-source Java library for personal data anonymisation and risk assessment;

pycanon, a Python library that provides standard metrics for anonymity; and *Anjana*, a Python library for anonymising sensitive data. Privacy assessment of tabular data typically involves two key metrics: k-anonymity and l-diversity, which characterise data similarity/indistinguishability. k-anonymity (Sweeney, 2002) is a widely used metric for anonymising tabular data. A dataset verifies k-anonymity for a certain value $k$ if every record (or participant represented in the dataset) is indistinguishable from at least *k-1* other records regarding the set of quasi-identifiers. l-diversity (Machanavajjhala et al., 2007) applies to sensitive personal data and measures the diversity of values in the sensitive attribute. Therefore, l-diversity is verified for a certain value of $l$ if, for each equivalence class in the dataset, the sensitive attribute takes at least $l$ different values.

Of notable interest for tabular data with quantitative variables are the Special Uniques Detection Algorithm (SUDA; Elliot et al., 2005) and the Personal Information Factor (PIF; CSIRO's Data, 2021), as those algorithms have been effectively utilised in governmental settings to enhance data privacy practices. SUDA is utilised by the UK Office for National Statistics, and the Australian government has utilised PIF to release COVID and Domestic abuse data. The Special Uniques Detection Algorithm (SUDA - Elliot et al., 2005) is a method for identifying records in a dataset that are unique with respect to specific combinations of variables, which can pose a disclosure risk (indistinguishability). SUDA examines all possible combinations of attributes, ranging from pairs of attributes to larger sets involving multiple attributes, to detect minimal sample uniques (MSU). Starting from smaller combinations and progressing to larger ones, SUDA considers the impact of each additional attribute on the dataset's uniqueness. In formal terms, SUDA is calculated and defined as follows: Be $m$ is the total number of variables. If we get a row with $k$ columns that have unique values within the entirety of the dataset, then we calculate SUDA as follows:

$$SUDA = (m - k)!$$

Let's take the example of a dataset consisting of 5 variables ($m = 5$), and we find that a given row with {Age: 98} and {Marital Status: widowed} is a unique combination, then SUDA = (5-2)! = 6. If age had been unique, not in combination with marital status, then SUDA = (5-1)! = 24. This means that SUDA assigns higher scores to smaller sample uniques, indicating that if a row has unique individual entries, the row is seen as less safe than if the row had a larger combination of unique samples. Importantly, if more sample uniques are found within the row, the sum of the sample uniques becomes the total SUDA score. The Personal Information Factor (PIF - CSIRO's Data, 2021) quantifies the level of information related to each individual across all variables and evaluates the associated risks of disclosure or misuse (information gain/loss). This is done using the Kullback–Leibler divergence, which assesses the information gain of a variable, given that an adversary knows everything about a subject in a dataset except for one thing. This divergence is calculated as follows:

$$D_{KL}(P|Q) = -\sum_{x \in X} p(x) \log q(x) + \sum_{x \in X} p(x) \log p(x)$$

*with* $p(x)$ *the prior for a given variable's distribution and* $q(x)$ *the posterior of the same variable's distribution conditioned on all other variables.*

For example, if we observe that the distribution of males and females is 0.5, but that a female race driver is 0.1, while a male race driver is 0.9, a significant information gain is disclosed by having the profession variable in the dataset, particularly for females. In this manner, PIF computes the cell information gain (CIG) across the entire dataset.

**Method**

Open-access BIDS datasets participants.tsv were analysed for privacy, aiming to identify which variables pose risks of identification in common brain imaging datasets and how this can be improved. As a secondary objective, privacy metrics were compared and correlated to better understand how they work and relate to one another in this context. Consistent with the OpenNeuro data usage agreement, we did not attempt to reidentify any subjects; we simply assessed the possible risk of reidentification.

*Datasets*

Datasets were pseudo-randomly selected from the OpenNeuro repository (Markiewicz et al., 2021) to encompass a wide range of use cases, including healthy participants, patients, adults, and children, as well as large and small sample sizes and metadata content, resulting in 6 datasets (Table 1, Aerts & Marinazzo, 2022; Gordon et al., 2023; A. T. Park et al., 2022; D. Park et al., 2024; Singh & Cavanagh, 2023; Snoek et al., 2021). This selection enables us to demonstrate how data privacy tools function in these contexts and to evaluate their relationships, alongside a qualitative analysis of the shared variables and associated risks. This does not reflect any particular judgment on the risk in those datasets. For full disclosure, each corresponding author of the datasets was contacted, and all reviewed the presented results and agreed with our assessment and discussion on variables and potential risks. Note also that while some subjects may be at risk in those datasets, no direct identification or obvious cross-linkage is possible, meaning that those datasets are acceptable from a privacy perspective. The goal here is to demonstrate how one can enhance privacy and identify across datasets, which variables require particular attention when releasing data.

| Name | ID | participants | sample size | metadata size |
|---|---|---|---|---|
| Amsterdam Open MRI Collection | ds003097 | Healthy adults<br>Age 22.8 ± 1.7<br>M/F = 0.9 | 928 | 27 |
| EEG: Alcohol imagery reinforcement learning task with light and heavy drinkers. | ds004515 | 26 Healthy adults<br>28 patients<br>Age 37.9 ± 10<br>M/F = 0.68 | 54 | 6 |
| The Dallas Lifespan Brain Study | ds004856 | Healthy adults<br>Age 58.3 ± 18<br>M/F = 0.6 | 464 | 35 |

| The Midnight Scan Club (MSC) dataset | [ds000224](https://openneuro.org/datasets/ds000224) | Healthy adults<br>Age 29.1 ± 3.3<br>M/F = 1 | 10 | 4 |
|---|---|---|---|---|
| Early stressful experiences are associated with reduced neural responses to naturalistic emotional and social content in children | ds004228 | Healthy children<br>Age 7.8 ± 1.7<br>M/F = 2.1 | 70 | 24 |
| Brain Tumor Connectomics Data | ds002080 | 10 Healthy adults<br>29 Patients<br>Age 57.8 ± 11.5<br>M/F = 0.8 | 39 | 50 |

*Table 1: Characteristics of selected datasets with name and Open Neuro ID. The participants column reports the type of participants, the mean age and standard deviation and the male/female ratio. The metadata size reported reflects the number of demographic variables (not the total number of variables in the participants.tsv, which could, for instance, include scanner-related parameters).*

*Metrics*

Open-access datasets metadata were analysed using the python MetaprivBIDS application (https://github.com/CPernet/metaprivBIDS), which applies the four metrics mentioned above (k-anonymity, l-diversity, SUDA scores, and PIF scores) and a new metric, k-global. While k-anonymity, k-global and l-diversity are computed directly within the app (counting 'uniques' - local unit tests), SUDA scores are obtained by importing the sdcMicro R library (https://cran.r-project.org/web/packages/sdcMicro/sdcMicro.pdf), while PIF is obtained from python the pif library (https://github.com/PIFtools/piflib). One advantage of using MetaprivBIDS over alternative libraries is that it provides a single point of entry for all those metrics and reads the tsv file along with the json sidecar file, helping identify variables and variable types (categorical vs continuous) and thus compute the right metrics automatically.

The k-global metric is a new metric developed for this application, because variables in neuroimaging datasets are not independent; it was therefore of interest to measure k-anonymity with respect to all other quasi-identifiers. A value of k-global can be thought of as its k-anonymity contribution for the entire dataset, and higher k-global values indicate variables that contribute more to uniqueness. The formal definition is given below (equation 1):

Let $n$ be the number of quasi-identifiers of a dataset $\Omega$, $U_i$ the number of unique values for the quasi-identifier $i$ ($\forall i \in \{1,..., n\}$), $U_n$ the number of unique combinations of the variables in $\Omega$ and $U_r$ the number of unique combinations for all variables but variable $i$. Then, we define the metric $K_i - global$ for the quasi-identifier $i$ as follows:

$$K_i \, global = \frac{U_n - U_r}{U_i} \quad \textbf{eq. 1}$$

Table 2 illustrates three cases to understand how k-anonymity, l-diversity and k-global relate to each other. Among six subjects for which sex and some disease information is recorded, one needs a combination that appears at least k-times, preventing a participant from being singled out (a precondition to identification) and needs to ensure enough diversity within each k-fold to avoid 'attackers' gaining information on a participant. For instance, if someone knows their female neighbour participated in a research study and there is only one female in the dataset, the attacker can check whether she is sick (see the left-hand side of the table). Even if there are two females, if the sampling is not diverse (i.e., both are from the same area and both are sick), the attacker will know whether her neighbour is sick (see the middle of the table). The k-global metric captures this well by equating the risks across variables (on the right-hand side of the table). In this table, the quasi-identifiers are Sex assigned at birth and Area (disease status being a sensitive attribute is not included, it counts toward l-diversity), leading to, for the 1st scenario (most left-hand side of the table), two unique rows for *all* quasi-identifiers (Female, City X) and (Male, Rural). If we remove the Sex assigned at birth variable, there is 1 unique row (Rural) in the reduced set, giving a K(sex) global of (2-1)/2 = 0.5. Only having enough people per group and diverse values can prevent privacy issues. Because k-global is a relative metric, it indicates how a variable weighs relative to others in the anonymity balance and is thus best used to detect subsets of variables that might need adjustments in their resolution (e.g., aggregating subgroups).

| k-anonymity 1<br>l-diversity 1<br>k-global (Sex=0.5, Area=0.67)<br>only 1 female and unique area | | | k-anonymity 1<br>l-diversity 1<br>k-global (Sex = 0.5, Area = 0.67)<br>at least 2 females but unique area | | | k-anonymity 2<br>l-diversity 2<br>k-global (Sex=0.5, Area=0.5)<br>at least 2 females and non-unique areas with diverse disease status | | |
|---|---|---|---|---|---|---|---|---|
| Sex assigned at birth | Area | Disease status | Sex assigned at birth | Area | Disease status | Sex assigned at birth | Area | Disease status |
| Female | City X | Yes | Female | City X | Yes | Female | City X | Yes |
| Male | Suburb | Yes | Female | City X | Yes | Female | City X | No |
| Male | Rural | Yes | Male | Rural | Yes | Male | Rural | Yes |
| Male | Suburb | No | Male | Suburb | Yes | Male | Rural | No |
| Male | City X | No | Male | City X | No | Male | City X | Yes |
| Male | City X | No | Male | City X | No | Male | City X | No |

*Table 2. Example of tabular data with only two variables: sex and disease status, illustrating how sampling and diversity are mandatory to avoid identifying or gaining knowledge on participants.*

*Detection of individuals at risk of identification:* While SUDA and PIF have the advantage of quantifying privacy, they lack absolute thresholds for detecting individual subjects. PIF creators recommend a value of 0.04 as a safe threshold for releasing open data, but also acknowledge that this is arbitrary. The MetaprivBIDS application, therefore, includes a robust outlier-detection method (MAD Median rule; Wilcox, 2012) to identify potentially problematic participants. This allows users to interact with a dataset and decide whether variables need to be removed or modified, and whether some subjects are more at risk than others.

## Analyses

### A. Dataset assessments

For each dataset, we (1) assess the privacy and (2) show how one can improve, where meaningful, the participants.tsv file. We report on variables that can be considered quasi-identifiers (variables that can be used to identify an individual through association with another variable) and those with sensitive attributes (data that should be protected under regulations or policies). Note that the selection of variables as having sensitive attributes in those datasets is based on the standard definition of special sensitive data observed across regulations (*DLA Piper Global Data Protection Laws of the World - World Map*): racial or ethnic origin, a person's sex life or sexual orientation, and data concerning health. Thus, we report on their L diversity along with average k-global anonymity. Next, while generally safe, we discuss which subjects are potentially at risk using outlier detection in SUDA and PIF scores. Lastly, and importantly, we demonstrate what can be done to improve privacy without altering meaningful associations with brain measurements a priori.

### B. Correlations among Metrics

Correlation analyses were conducted to understand the relationships among these metrics and their contributions to risk understanding and identification. We computed Spearman correlations between SUDA and PIF scores for each dataset, using each subject's scores. In addition, we also conducted a global analysis across all subjects and datasets. Since SUDA represents the risk associated with uniqueness relative to all other variables, and PIF represents the risk associated with conditioning on all other variables, we hypothesised that the two variables would be correlated. We computed Spearman correlations between k-global (the contribution of a variable to the dataset's uniqueness, k-anonymity), the SUDA Field scores (a SUDA score at the variable level), and Field Information Gain (the sum of PIF scores per variable). Since SUDA is essentially a raw (i.e., participant) measurement, no association with k-global was hypothesised. By contrast, because PIF accounts for variable distributions, we hypothesised that some associations would be observed with k-global.

C. Computational reproducibility

The archived version of metaprivBIDS used for the analysis is available at Zenodo @https://doi.org/10.5281/zenodo.17150814 and JUPYTER notebooks were used for each dataset privacy assessment and correlation analysis. This allowed anyone to reproduce our analyses and learn how to call the metaprivBIDS functions. These are accessible @ https://github.com/CPernet/OpenNeuro_MetaPrivAssessment.

## Results

After searching and selecting datasets on OpenNeuro, we identified two problematic datasets (out of eight) that contained a unique combination of variables that could identify a participant if we had performed cross-linkage, which represents 2 individuals out of 3662 included across datasets. Those datasets have been altered from OpenNeuro records and are not discussed here for privacy and ethical reasons. This is perhaps the most significant result: we identified datasets that needed changes, thereby improving participants' privacy. The six datasets below, on the other hand, pose only minimal risks but are a good illustration of issues that must be addressed when curating datasets.

### *1. Amsterdam Open MRI Collection*

*Variable classification*: We identified 10 out of 27 variables as quasi-identifiers (see Table 3), with two variables seen as sensitive: sexual orientation and IST Intelligence (l-diversity of 1). Here, we chose to include sexual orientation in a k-anonymity computation because, for low/high levels, individuals likely live with or have a partner of the same/different gender, which is a visible attribute and, therefore, needs to be homogenised. The k-global anonymity for these variables ranges from 0.3 to 42.7, with a mean value of 17.7, which is significantly lower than the other variables (t(dt)=4.25, p=0.0003).

| Variables | Description | Unique Values | k-global | Sensitive |
|---|---|---|---|---|
| Education level | Level of education as defined by the Central Bureau van de Statistiek, CBS. | 3 | 42.7 | no |
| background SES | Socio-economic status (based on parents' income and educational level). | 9 | 41.8 | no |
| Sex | Gender of the participant at birth | 2 | 34.0 | no |
| Handedness | Preferred hand used. | 2 | 18.5 | no |
| Age | Age reported with one quantile precision. | 27 | 18.3 | no |

| gender identity (Male/Female rating - 2 variables) | To what extent the subject identifies as male or female. | 7/7 | 2.4/1.0 | no |
|---|---|---|---|---|
| sexual attraction (Male/Female rating - 2 variables) | To what degree the subject feels attracted to males or females. | 7/7 | 0.3/0.3 | yes |
| IST Intelligence Total | Intelligence Structure Test global score | 182 | NA | yes |

*Table 3: Variables for risk identification in the Amsterdam Open MRI Collection study.*

Correlations and outliers: Using SUDA and PIF scores, 440 outliers were identified among 928 participants, with six outliers in common. Since the Spearman correlation between SUDA and PIF is high (r=0.76, p<0.001), it makes sense to consider outliers in common. Those metrics did not correlate with k-global (k-global-SUDA: r = -0.54, p = 0.13; k-global-PIF: r = -0.28, p = 0.46). These six outliers can be viewed as at-risk individuals with a unique combination of variable values, as they have higher risk scores than other individuals in the dataset.

Improving privacy: sub-0339 is a 22.8-year-old female with a high educational and socioeconomic status (level 6), self-identifying as both female and male (4.0/3.0) and a higher attraction to the same sex (F 6.0/M 3.0), which is identified as an outlier, and thus at higher risk. This higher risk stems from both the sexual attraction and gender identity attributes of being in a minority group. Privacy can be improved by examining the prevalence of such cases (i.e. diversity). If sexual attraction is not scientifically relevant, then removing the variable will improve anonymity. If, on the contrary, that is of interest, one can act on another variable, e.g. background SES. This variable has a high k-global value, indicating that it contributes a lot to 'uniqueness' in the dataset. Simply binning values 2 and 2.5, because they cluster with education level, brings the total unique combinations down from 716 to 452 (-36.9%), so no common outliers can be detected (see k_global.ipynb for a detailed analysis).

### *2. Brain Tumor Connectomics Data*

Variable classification: We identified 7 out of 50 variables as quasi-identifiers (see Table 10), with one sensitive attribute (tumor type & grade) with l-diversity = 1. Again, these sensitive attributes are not directly visible and, therefore, included in the computation of l-diversity. The k-global anonymity for these variables ranges from 0.0 to 0.4, with a mean value of 0.07, significantly higher than the other variables (t(dt)=3.35, p=0.002)).

| **Variables** | **Description** | **Unique Values** | **k-global** | **sensitive** |
|---|---|---|---|---|
| Education level (1-9) | International Standard Classification of Education | 5 | 0.4 | no |
| age | Age of participant in years. | 22 | 0.1 | no |
| height(cm)(pre-op) | Height of patient in cm. | 18 | 0.0 | no |
| handedness | handedness measured using the Edinburgh Handedness Inventory | 6 | 0.0 | no |
| Employment | Current working status of the patient. | 5 | 0.0 | no |
| Marital status | Self reported Marital status from the patient. | 4 | 0.0 | no |
| sex | Gender of patient. | 2 | 0.0 | no |
| tumor type & grade | Tumor grade and type. | 9 | NA | yes |

*Table 10: Variables for risk identification in the Brain Tumor Connectomics study.*

Correlations and outliers: Using SUDA and PIF scores ($r=0.80$ $p<0.001$), no outliers were found. k-global also did not correlate with SUDA and PIF ($r=0.58$ $p=0.17$, $r=0.58$ $p=0.71$).

Improving privacy: The dataset is at very low risk of identification, but improvement is possible by generalising the Marital status variable, combining “Divorced” & “widowed” into “Single”, and combining “Married” & “Cohabitating” into “In Relationship”. This homogenises data without decreasing utility. Similarly, removing height would further enhance data privacy risk, thereby preventing a background attack. For instance, an adversary who knows her colleague or neighbour is part of the study could try to identify her. Using employment and marital status (especially after merging) is not helpful, as many entries are identical, but height is essential, as only one female is above the average (180 cm). Since such a variable is irrelevant to the imaging data, it provides unnecessary risk.

### ***3. EEG: Alcohol imagery reinforcement learning task with light and heavy drinker participants.***

Variable classification: We classified all six variables as quasi-identifiers (see Table 4), with three variables (BDI, AUDIT, and GROUP) being sensitive, each with an l-diversity of 1. Since there are no visible attributes, none were included in the k-anonymity computation.

| Variables | Description | Unique Values | k-global | Sensitive |
|---|---|---|---|---|
| sex | Gender of the participants. | 2 | 2.0 | no |
| age | Age in years of participants. | 30 | 1.1 | no |
| group | Alcoholic vs. non-alcoholic | 2 | NA | yes |
| BDI | Beck Depression Inventory self-report. | 19 | NA | yes |
| AUDIT | Screening tool for alcohol use disorders. | 17 | NA | yes |
| education | Categorical Education of participants. | 12 | 1.1 | no |

*Table 4: Variables for risk identification EEG: Alcohol imagery reinforcement learning task with light and heavy drinker participants study.*

Correlations and outliers: Using SUDA and PIF scores ($r=0.68$ $p<0.001$), four outliers were found with PIF.

Improving privacy: This dataset is safe, given the low number of quasi-identifiers and the fact that the education variable is numeric with no reference categories. While the risk is very low, we here illustrate that it is never null. For instance, in a background attack where an adversary knows a participant's involvement in the study and their age, re-identification is possible. For example, participant sub-031 is the only female participant aged 25. In the event of such an attack, an adversary would then discover that the participant had endorsed problematic alcohol abuse. To prevent such an attack, a potential improvement could be to add noise to the age variable. After doing so, we can achieve an l-diversity score of 2 for the GROUP variable, 11 for AUDIT and 12 for BDI.

***4. Early stressful experiences are associated with reduced neural responses to naturalistic emotional and social content in children***

Variable classification: We identified 7 out of 24 variables as quasi-identifiers (see Table 8), with one sensitive attribute (Race). The k-global anonymity for these variables ranges from 0.0 to 1.8, with a mean value of 0.46, significantly higher than the other variables ($t(dt)=2.79$, $p=0.01$).

| Variables | Description | Unique Values | k-global | sensitive |
|---|---|---|---|---|
| age scan | The child's age at the time of the | 8 | 1.8 | no |

| | MRI. | | | |
|---|---|---|---|---|
| parent 1 edu | The highest degree the parent 1 has achieved. | 6 | 1.0 | no |
| Race | Race of the child. | 6 | 0.4 | yes |
| income median | Median of the parents' income bracket. | 10 | 0.0 | no |
| parent 2 edu | The highest degree the parent 2 has achieved. | 6 | 0.0 | no |
| male | The child's gender, as reported by the parent. | 2 | 0.0 | no |
| income rank | Total combined family income for the past 12 months. | 11 | 0.0 | no |

*Table 8: Variables for risk identification in the Early Stressful Experiences in Children study.*

*Correlations and outliers*: Using SUDA and PIF scores ($r=0.82$ $p<0.001$), two outliers were found with PIF. k-global did not correlate with SUDA and PIF scores ($r=0.39$ $p=0.38$, $r=-0.02$ $p=0.97$).

*Improving privacy*: Overall, the dataset doesn’t reveal any compromising information to an adversary, except for the participant's race. One possible issue could be with sub-08, a 9-year-old black male living in a household with a total income bracket of $5,000 to $11,999. The parents' highest educational degree obtained is a bachelor's degree. The flagging of this particular participant stems from the minority group of black/African Americans within the dataset, with only 14 participants out of 70 being of this racial background, highlighting the need for careful data sampling during data acquisition. Furthermore, the reported household income is at the lower end of the scale, making the entry stand out in the dataset. Income here has not been chosen as a sensitive attribute, as it is reported in the context of the whole household. However, to make the dataset safer, in the sense of making it more difficult to breach, one could remove the income values/brackets if deemed not relevant to imaging or substitute for broader categories (e.g. above/below US median income).

***5. The Dallas Lifespan Brain Study***

*Variable classification*: We identified 8 out of 35 variables as quasi-identifiers (see Table 5), with two variables (Race and ethnicity) being sensitive. We include the two sensitive variables in the computation of k-anonymity as they are visible attributes; hence, diversifying them would not help prevent re-identification. The k-global anonymity for these variables ranges from 0.0 to 1.6, with a mean value of 0.4, significantly lower than the other variables (t(dt)=3.26, p=0.003).

| Variables | Description | Unique Values | k-global | sensitive |
|---|---|---|---|---|
| Race | Race that the participant self-identifies with | 7 | 0.0 | Yes |
| EduYrsEstCap | Years of education, capped by degree | 16 | 0.0 | no |
| HandednessScore | Average score of participant hand preference on the Edinburgh Handedness Inventory | 21 | 0.3 | no |
| EduComp | Highest level of education completed | 7 | 0.0 | no |
| Sex | Participant's biological sex | 2 | 0.0 | no |
| Ethnicity | Ethnicity that the participant self-identifies with | 2 | 0.0 | Yes |
| Height W1 | Participant height at wave 1 | 30 | 1.3 | NA |
| AgeMRI W1 | Age of the participant at the wave 1 MRI scan | 67 | 1.6 | NA |

*Table 5: Variables for risk identification for The Dallas Lifespan Brain Study.*

Correlations and outliers: Using SUDA and PIF scores (r=0.57 p<0.0001), 14 outliers were found with PIF and none with SUDA. k-global also correlated with SUDA (r=0.73 p=0.03) and PIF (r=0.80 p=0.03).

Improving privacy: the risk is overall low, though a few outliers remain. Improvement can be achieved by reducing the resolution of the height and age variables, thereby preventing a possible background attack. One way to improve anonymity is to use body mass index instead of height, as height is unlikely to be a relevant feature for brain imaging (while BMI can be). Here, removing height reduces the number of unique rows from 350 to 159, highlighting the importance of eliminating continuous variables when applicable.

***6. The Midnight Scan Club (MSC) dataset***

Variable classification: All four variables used in the dataset are quasi-identifiers (see Table 7), with no sensitive attributes.

| Variables | Description | Unique Values | k-global | sensitive |
|---|---|---|---|---|
| education years | Number of total years of education. | 7 | 0.3 | No |
| age | Participants age. | 7 | 0.3 | No |
| education degree | Highest educational degree obtained. | 4 | 0.0 | No |
| gender | Birth gender. | 2 | 0.0 | No |

*Table 7: Variables for risk identification in the Midnight Scan Club dataset*

*Correlations and outliers*: Using SUDA and PIF scores (r=0.69 p=0.02), one outlier was found with PIF. The identified participant is sub-MSC02, a 34-year-old male with a doctorate who has studied for 28 years.

*Improving privacy*: Despite this participant being flagged, the dataset contains no sensitive attributes and has a minimal number of quasi-identifiers, making it a low-risk.

**Discussion**

While much work has focused on brain imaging identifiability (Jwa et al., 2023), more attention needs to be paid to demographic and clinical data. Those data are an easy target because they require little expertise to analyse and no specific or proprietary software (Rocher et al., 2019). If privacy is not checked, they can thus lead to reidentification. Here, we performed such privacy checks using state-of-the-art privacy tools and observed that risks are real, with 2 people (0.05%) potentially identifiable among all 3662 participants across 8 datasets. From the six datasets presented, we further identify which types of variables are most likely to drive reidentification risk, providing practical guidance on what requires particular caution prior to data release.

More broadly, these findings highlight the diversity of strategies currently used to address privacy risks in neuroimaging research. Concerns around privacy, governance, and regulatory constraints have motivated a growing ecosystem of approaches to limit direct data exposure, including federated and privacy-preserving learning frameworks, where data remains local and only derived parameters are shared. These approaches are particularly relevant in settings where data centralisation is infeasible. By contrast, the present work addresses a complementary but distinct point in the data-sharing landscape: situations in which datasets are intended for open or controlled public release, and where privacy risks must therefore be assessed and mitigated prior to dissemination. Rather than replacing federated or distributed analyses, the proposed framework provides an additional, orthogonal tool that supports informed decision-making about data release by quantitatively characterising privacy vulnerabilities in tabular metadata. Together, such approaches reflect a broader toolbox for addressing privacy concerns across the full spectrum of data access and sharing models.

*Demographic data*

Many variables used regularly in neuroimaging research are quasi-identifiers: age, sex assigned at birth, handedness, educational level, etc. These are essential characteristics linked to brain features (Brain Imaging in Normal Subjects Expert Working Group et al., 2017) and must, therefore, be present in datasets. From a scientific perspective, having a diverse representation (i.e., a good sampling across categories) is essential to capture the range of human variations (high utility). From a privacy perspective, it is critical to avoid reidentification of individuals, which is maximised by homogeneous sampling (if every participant is from a single group, no one can be singled out). There is, therefore, an opposition between scientific utility and privacy. The practical solution is to ensure enough participants in each subgroup (ideally in the same proportion) and, if not, consider grouping values within variables. Our analyses revealed that variables posing the highest risks across datasets were demographic (such as age and visible attributes like race or height) and variables that could geolocate individuals.

Age is a defining feature of brain morphology and function, but it is also a variable that can lead to reidentifying individuals. Since age is present in all datasets, we urge researchers to pay closer attention to it. Resolution (i.e. how fine-grained age is reported) should be considered first. In studies of young children, months are helpful, but years, or even biennia or lustra, are likely enough in the general adult population. Second, extreme values need to be examined, especially in ageing studies. HIPAA, for instance, does not allow sharing participants' exact ages above 89. Finally, because age is, in many smaller studies, a set of unique values, combining it with other variables increases the number of unique values. One option is to bin participants into subgroups based on the distribution's standard deviation, leading to fewer unique values. Another option is to add noise, e.g., by drawing values from a Gaussian distribution or, if extreme age values are observed, from a Laplacian distribution, thereby making direct reidentification less likely. It is important that researchers consider the utility of the measurement for decision-making (Brummet et al., 2021). Some approaches, like binning, might be good enough to keep utility (e.g. regressing out age) while others might alter it (e.g. adding noise while wanting to correlate age with brain features),

Visible attributes are variables particularly sensitive to background attacks, in which an adversary has prior knowledge about an individual, such as knowing that someone participated in a research study. In such a scenario, a lack of diversity (i.e., a unique case) inevitably leads to individuals being identified and the attacker gaining information about them. An example of this could be a unique individual from Brain Tumour Connectomics Data Research. Unless measures are taken to make sure that there is homogeneity in the quasi-identifiers, meaning multiple people have the same marital status, education, etc., risk from a background attack is hard to prevent.

While there may be valid scientific reasons to share geolocation information (for instance, areas of exposure to environmental factors), such information should be handled with extreme care. HIPAA, for example, has strict guidelines on handling geographic data, specifying that all geographic subdivisions smaller than a state, like street addresses, cities, and zip codes, must be restricted. It allows the initial three digits of a zip code to be used only if they represent an area with more than 20,000 people. If the population is smaller, the zip code must be changed to '000' to prevent identification (Office for Civil Rights, 2008). However, research has highlighted that even with these protections, other quasi-identifiers in the data, when combined, can still potentially narrow down individuals within these larger groups (Liu et al., 2022). This suggests that relying solely on HIPAA's geographic masking is a naive approach to data privacy, particularly when multiple data points can be linked to deduce personal identities, such as income or socioeconomic status. In that regard, geolocating variables are particularly susceptible to linkage attacks, in which an attacker uses external sources of information. A crucial part of data privacy that is often overlooked is that it should not be viewed solely at the individual level, but in the context of other variables, starting with k-anonymity. Sweeney (2000) showed that the combination of 5-digit ZIP, gender, and date of birth could uniquely identify 87% of the U.S. population. Additionally, 53% could be uniquely identified using place (city, town, or municipality), gender, and date of birth. At the county level, the combination of county, gender, and date of birth uniquely identifies 18% of the population, illustrating that very few characteristics are needed to identify a person. This demonstrates that all variables must be inspected together; hence, metrics such as k-global, SUDA, and PIF are needed to help identify issues. Without properly accounting for unique combinations across all variables with geolocation information, re-identification of individuals is likely (Douriez et al., 2016; Lavrenovs & Podins, 2016; Sweeney, 2018; Yoo et al., 2018).

*Clinical Data*

Clinical data falls under the special category of sensitive data under many legislations (see introduction), along with (depending on regulations) racial or ethnic origin, political opinions, religious or philosophical beliefs; trade-union membership, genetic data, biometric data used to identify a human being (iris, fingerprints); and data concerning a person's sex life or sexual orientation. This does not mean we cannot share that data; they simply need to be lawfully processed in accordance with the relevant legislation, and always after removing all direct identifiers and assigning a pseudo-ID. It also applies to identifiers included in brain imaging data, which is typically automatically handled during the DICOM-to-Nifti conversion. Still, special care should be taken if one is to share the source data. Importantly, none of the clinical data (IQ, Tumour status, Depression scores, and alcohol use) were flagged as a source of re-identifiability in our analyses. A more typical issue that can arise with clinical data is having a unique or small number of participants in a demographic variable (i.e., l-diversity), which may reveal sensitive clinical information about those subjects. Even so, when curating datasets, despite good K and L values, it remains possible that individuals are at risk.

*In-sample risk*

One might wonder how likely a participant is at risk in large datasets such as the UK Biobank (Littlejohns et al., 2020) or the ABCD study (Casey et al., 2018). As demonstrated in Sweeney (2000), only a few attributes are sufficient for identification, and Rocher et al. (2019) estimated that with 15 attributes, 99.98% of all Americans could be identified, demonstrating that the total number of participants is irrelevant. The risk of re-identification is actually a function of both the dataset and the population it comes from (Shlomo & Skinner, 2022). In short, while the in-sample uniqueness may be reduced with a larger dataset, it does not necessarily lessen the risk of identification if uniqueness is also high in the reference population. For instance, in the Brain Tumor Connectomics Data (ds002080), participants are not drawn from the general population, and since brain tumour is overall a rare condition, the risk of identification is de facto increased. A background attack becomes easier, and having fine-grained variables increases the probabilistic risk (if not the individual risk) of identification because an attacker can more easily identify someone from that reference population.

*Privacy Metrics*

There are many methods and metrics for measuring privacy. The analysis presented here is restricted to just five of them. They do, however, represent commonly used approaches, not just in research settings but also in global information applications by government officials to release sensitive data. In our analysis, common PIF-SUDA outliers were often effective at identifying issues (e.g. Amsterdam Open MRI Collection), although in smaller datasets, only PIF flagged outlying data points. Still, those metrics can also flag a unique individual with no risk (e.g. The Midnight Scan Club dataset) because these metrics have high sensitivity but low specificity. Common outliers can generally be expected, as those metrics were correlated within and across datasets (r=0.75, CI [0.741 0.758]; Figure 1, top), as hypothesised.

While in some cases, no SUDA-PIF outliers corresponded to no risk, this was not always the case. The new metric developed here, k-global, tries to capture the significance of individual variables in determining a dataset's k-anonymity by removing each variable and observing the resultant changes in the number of unique rows, normalised by the number of unique values within the removed variable, which helps to assess each variable's proportional impact on privacy in the context of k-anonymity. Acting on variables with high k-global values can reduce a dataset's uniqueness and thus decrease overall risk. While this approach highlighted the roles of different variables, it might overlook the subtler, non-linear interactions among them, which are better captured with SUDA and PIF. We hypothesised that k-global and PIF information gain were related, since they both measure the relative contribution of a variable to the set, but no correlation was observed, except in the Dallas Lifespan Brain Study (Figure 1). Practically, it means that k-global, SUDA field score, and PIF field information gain provide different variable-level information. This highlights that metrics can help find issues, but are not the definitive solution. Our analysis suggests that it is beneficial to consider quantifiable methods for assessing privacy risks, but also that metrics are not a substitute for the human-defined importance of individual variables.

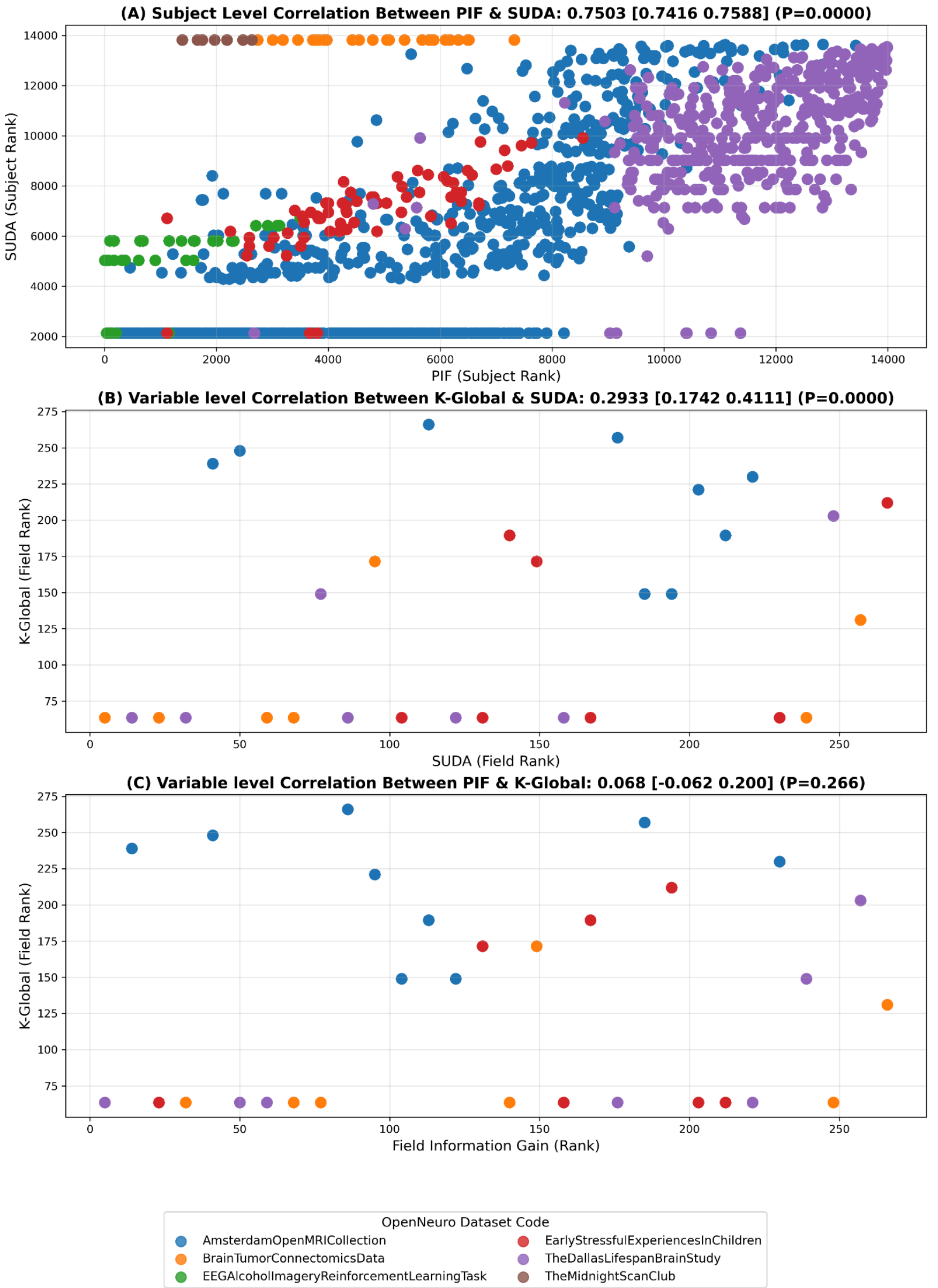

*Figure 1: Spearman correlation analyses. At the top is the analysis for all subjects SUDA and PIF values across studies. In the middle and bottom are variable-level correlations for k-global vs. SUDA field score and PIF field information gain. Results are reported with a 95% percentile bootstrap confidence interval.*

In our opinion, investigators who define the data-acquisition protocols and researchers deeply involved in data collection and analysis are best able to curate the data. Expertise on privacy may be available in libraries, databases, and repositories, but (1) only a good understanding of the variable at hand can help determine which variables are sensitive and (2) a knowledge of the participants' demographics and how to coarsen variable resolution or add noise without reducing utility. *MetaprivBIDS* aims to fill a gap by providing brain imagers (psychologists, medical doctors, biologists) a single entry point to compute all metrics and help resolve problems. The software application offers a solution designed explicitly for neuroimaging to evaluate variables stored in the participant.tsv/json files. The archived version used for the analysis presented here is available at Zenodo @https://doi.org/10.5281/zenodo.17150814

## Conclusion

In this work, we demonstrate that privacy metrics help identify problematic variables (typically demographics or geolocating variables with too fine-grained resolutions) and detect participants with unique combinations that may put them at risk of re-identification. The solution to such a privacy risk is sometimes to remove these variables or participants, but more often, changing variable resolution or adding noise is sufficient. Given the tension between scientific utility (maximally diverse individuals to represent the population) and privacy (maximally homogeneous characteristics), it is worth considering the options available to researchers to achieve transparency. After analysing demographic and clinical data and modifying them to ensure privacy, we suggest that researchers indicate which variables were not released, document variables' initial resolution vs. the released version (for instance, in the README file of BIDS datasets), and indicate where noise is added. Another solution is to mask some values for participants at risk, thereby allowing the release of all imaging data while preserving privacy.

## Ethic

All datasets used have open licences enabling reuse

## Data and Code Availability

All data are available on OpenNeuro. The library metapivBIDS is available @https://github.com/CPernet/metaprivBIDS under MIT licence. The Code used for the analyses is available as a series of notebook @https://github.com/CPernet/OpenNeuro_MetaPrivAssessment

## Author Contributions

E Kibsgaard: Formal Analysis, Investigation, Methodology, Writing. CJ Markiewicz: Investigation, Review & Editing. J Sainz Pardo: Methodology, Review & Editing.. AS Jwa, D Rodriguez Gonzalez, RA. Poldrack: Review & Editing. CR. Pernet: Conceptualisation, Formal Analysis, Methodology, Funding Acquisition, Writing.

## Declaration of Competing Interests

We declare no conflict of interest.

## Funding

This work is supported by the European Union's Horizon Research and Innovation programme (SIESTA: Grant agreement 101131957).